# Depolarization diagrams for circularly polarized light scattering for biological particle monitoring

Nozomi Nishizawa[1*], Asato Esumi[1] and Yukito Ganko[1]

[1]Department of Physics, School of Science, Kitasato University, 1-15-1 Kitazato, Sagamihara, Kanagawa 252-0373, Japan.
*nishizawa.nozomi@kitasato-u.ac.jp

**ABSTRACT**

**Significance:** The depolarization of circularly polarized light caused by scattering in turbid media reveals structural information about the dispersed particles, such as their size, density, and distribution, which is useful for investigating the state of biological tissue. However, the correlation between depolarization strength and tissue parameters is unclear.

**Aim:** We aimed to examine the generalized correlations of depolarization strength with the particle size and wavelength, yielding depolarization diagrams.

**Approach:** The correlation between depolarization intensity and size parameter were examined for single and multiple scattering using the Monte Carlo simulation method. Expanding the wavelength width allows us to obtain depolarization distribution diagrams as functions of wavelength and particle diameter for reflection and transparent geometries.

**Results:** Circularly polarized light suffers intensive depolarization in a single scattering against particles of various specific sizes for its wavelength, which becomes more noticeable in the multiple scattering regime.

**Conclusions:** The depolarization diagrams with particle size and wavelength as independent variables were obtained, which are particularly helpful for investigating the feasibility of various particle-monitoring methods. Based on the obtained diagrams, several applications have been proposed, including blood cell monitoring, early embryogenesis, and antigen-antibody interactions.

---

* Address all correspondence to : Nozomi Nishizawa, 1-15-1, Kitazato, Sagamihara, Kanagawa, 252-0373, Japan. Tel: +81-42-778-9540; E-mail:nishizawa.nozomi@kitasato-u.ac.jp

# 1 Introduction

Light scattering phenomena by fine particles is classified based on the length relationship between the wavelength and particle diameter [1, 2]. When the incident light's wavelength is significantly larger than the particle diameter (i.e., in the Rayleigh scattering regime), the incident light excites a dipole on the particle. The excited single dipole oscillates with the same periodicity as the incident light, emitting secondary light in the form of scattered light. The intensity distribution of scattered light shows almost $\cos^2 \varphi$ to the scattering angle $\varphi$, which is nearly symmetric distribution for forward and backward scattering because the excited dipole can be regarded as a single. The depolarization strengths of $\varphi$ are also symmetric. Single scattering slightly rotates the incident light's polarized plane, resulting in depolarization to linearly polarized light (LPL) but not circularly polarized light (CPL). Alternatively, backscattering serves as an optical mirror for CPL, flipping its helicity from right to left and vice versa. For multiple scattering, the helicity flip significantly contributes to the CPL's depolarization. Eventually, the CPL and LPL are depolarized almost in the Rayleigh regime [3, 4].

In contrast, shorter wavelengths or larger particle sizes cause drastic changes in these symmetric behaviors, as explained by the Mie scattering theory [2]. In the Mie scattering regime, incident light excites multiple dipoles within a particle [5]. As they oscillate and emit secondary light with slight delays, forward scattering increases predominantly over backscattering. As a result, CPL is more persistent than LPL because less backscattered light causes the depolarization of CPL [3]. The distributions of intensity and depolarization by Mie scattering are asymmetrical and depend strongly on the particle size and wavelength. The spatial asymmetric patterns of Mie scattering are used in laser diffraction/scattering analysis to measure the geometrical dimensions of fine particles [6]. Thus, CPL scattering in the Mie regime has two application characteristics: CPL has greater persistence than LPL for multiple scattering, and the scattered light possesses a circular polarization corresponding to the scatterer's particle size [7, 8]. Scattering and propagation of CPL within from turbid disperse medium have been investigated by a combination of Jones and Stokes-Mueller formalisms in Monte Carl (MC) modeling approach to explore the fundamentals of CPL depolarization, which can be applied to practical uses in a wide range of fields [9–11].

Applications using these characteristics have also been studied. One such application is the detection of cancer distribution [12–18]. A large number of cell nuclei in biological tissues work as Mie scatterers for light ranging from visible to near-infrared light. Cells in most digestive cancers have enlarged or distorted cell nuclei; therefore, cancerous tissue can be identified by a difference in the circular polarizations of scattered light. Near-infrared light can penetrate biotissues to depths of several millimeters, and the circular polarizations of the incident light cannot be completely depolarized by hundreds of scattering events during the reciprocating travel of light to depth. Therefore, when the scattering volume with penetration depth includes some cancerous cells in normal tissue, depolarization is associated with the ratio of cancerous to normal cells. These characteristics can be used to estimate the distribution of cancerous cells. When cancerous tissues generates in the surface layer, the degree of cancer progression could be estimated from the variation of circular polarization with scattering angles [8, 18]. And, when a small number of cancerous cells mixed and diffused in healthy tissue, the larger scatterers (cancerous cell nuclei) in the scattering volume would bring a change on the polarization of resultant circular polarization. This indicates the probability of detecting a diffuse cancer such as early-staged scirrhous stomach cancer. Compared with the existing *in vivo* diagnostic techniques with endoscope, such as narrow band imaging [19], the CPL scattering technique has higher depth resolution but less in-plane resolution. The detected polarization values are averaged values of light scattered from a circular region centered at the incident point with a radius of approximately 1 mm. Therefore, the in-plane resolution of this technique is 1 mm at most. The less in-plane resolutions would be effectively utilized in the estimation of cancer progression and the detection of diffuse cancer.

For cancer detection and estimation using CPL scattering, light with wavelengths of approximately 600 and 950 nm is used [18]. These wavelengths show the opposite changes in depolarization of cancerous and normal cells; scattering depolarization of light of 600 nm for normal cell nuclei (~6 μm) exhibits stronger depolarization than that for cancerous cell nuclei (~11 μm), while light of 950 nm shows the direct opposite trend. A pair of wavelengths exhibiting opposite responses clarifies the differences between particles without being disturbed by other scattering and depolarizing factors. Similarly, one can determine the existence, density, distribution, and successive changes of the intended particle in dispersive materials by selecting the appropriate wavelengths corresponding to the radius of the intended particle and investigating and comparing the depolarization of the scattered CPL. Even in the case of a complicated medium in which particles of various sizes intermingle, the state of each particle can be investigated individually, unless the optimum wavelengths overlap.

In this study, we investigated the dependence of the depolarization strength on the size parameter using MC simulations. The size parameter is defined as the ratio between the wavelength of incident CPL and the particle diameter. Incident CPL photons undergo multiple scatterings in a turbid medium, in which the particles diffuse at a fixed density. Subsequently, the scattered light is detected on the same and the other sides, which are referred to as the reflection and transparent geometries, respectively. Finally, applications are proposed using the simulation results.

## 2 Theory

Polarized light can be decomposed into two components that oscillate coherently and perpendicularly. The polarization state of light of transversal electromagnetic waves is typically described by a four-element vector known as the Stokes vector $S$, expressed by the equation $S = (S_0, S_1, S_2, S_3)^T$, where $S_0$, $S_1$, $S_2$ and $S_3$ are Stokes parameters [20]. $S_0$ is the total light intensity, $S_1$ quantifies the fraction of linear polarization parallel to the reference plane, $S_2$ provides the proportion of linear polarization at 45° with respect to the reference plane, and $S_3$ quantifies the fraction of right-handed circular polarization.

The relationship between the Stokes parameters of the incident and scattered light in scattering theory [2, 21] can be expressed using the Mueller matrix as follows:

$$\begin{pmatrix} S_0' \\ S_1' \\ S_2' \\ S_3' \end{pmatrix} = \begin{pmatrix} M_{11} & M_{12} & 0 & 0 \\ M_{12} & M_{11} & 0 & 0 \\ 0 & 0 & M_{33} & M_{34} \\ 0 & 0 & -M_{34} & M_{33} \end{pmatrix} \begin{pmatrix} S_0 \\ S_1 \\ S_2 \\ S_3 \end{pmatrix}. \quad (1)$$

Each element of the Mueller matrix can be expressed as follows:

$$M_{11} = \frac{1}{2}(|M_2|^2 + |M_1|^2), \qquad M_{12} = \frac{1}{2}(|M_2|^2 - |M_1|^2),$$

$$M_{33} = \frac{1}{2}(M_2^*M_1 + M_2M_1^*), \qquad M_{34} = \frac{1}{2}(M_2^*M_1 - M_2M_1^*), \quad (2)$$

where

$$M_1 = \sum_n \frac{2n+1}{n(n+1)}(a_n\pi_n + b_n\tau_n),$$

$$M_2 = \sum_n \frac{2n+1}{n(n+1)}(a_n\tau_n + b_n\pi_n) \quad (3)$$

where the angle-dependent functions $\pi_n$, $\tau_n$, using Legendre functions $P_n^l$, are defined as follows:

$$\pi_n = \frac{P_n^l}{\sin\theta}, \quad \tau_n = \frac{dP_n^l}{d\theta}. \tag{4}$$

And $a_n$ and $b_n$ are the Mie coefficients expressed as follows:

$$a_n = \frac{m\psi_n(mx)\psi_n'(x) - \psi_n(x)\psi_n'(mx)}{m\psi_n(mx)\xi_n'(x) - \xi_n(x)\psi_n'(mx)}, \tag{5}$$

$$b_n = \frac{\psi_n(mx)\psi_n'(x) - m\psi_n(x)\psi_n'(mx)}{\psi_n(mx)\xi_n'(x) - m\xi_n(x)\psi_n'(mx)}. \tag{6}$$

In the equations of the Mie coefficients, $\psi_n(x)$ and $\xi_n(x)$ are Ricatti-Bessel functions. The relative refractive index $m$ is expressed as follows:

$$m = \frac{n_{particle}}{n_{medium}}. \tag{7}$$

$n_{particle}$ and $n_{medium}$ denote the refractive indices of the particles and medium, respectively. $x$ is the size parameter of Mie theory equal to

$$x = \frac{\pi a}{\left(\frac{\lambda}{n_{medium}}\right)} \tag{8}$$

where $a$ denotes the diameter of particle and $\lambda$ denotes the wavelength of light. Finally, the polarization states of the scattered light, $S_1$, $S_2$, and $S_3$, depend on the parameters $m$ and $x$.

## 3  Calculation methods

In this study, the polarization-light MC algorithm developed by Ramella-Raman *et al*. [22] was used for the depolarization calculations of scattered light. This MC algorithm is known as "meridian plane MC" because the polarization can be described by the Stokes vector $S$ with respect to the meridian planes, which is determined by the light propagation direction and a particular axis. In the meridian-plane MC algorithm, the scattering and azimuth angle of a single scattering event are selected using the rejection method [23], which generates random variables with a distribution in the Mie scattering process described in the previous section. The propagation direction and Stokes vector of the scattered light were calculated. First, the angular dependences of $S_0$ and $S_3$ components of the scattered light were calculated for a single-scattering event. The $S_0$ divided by total photon number provides the scattering probability. Subsequently, the resultant $S_0$ and $S_3$ distributions were introduced into standard optical MC programs, yielding a polarized MC program.

The polarization states of the CPL after multiple scattering events were calculated using two optical geometries: reflection and transparent configurations, as shown in Figure 1 (a) and (b), respectively. The scattering media are turbid aqueous mediums with dispersing particles with a diameter $a$ whose density and dispersions are expressed optically by the absorption coefficient ($\mu_a$) and scattering coefficient ($\mu_s$). In this study, these optical parameters were fixed that $\mu_a = 0.10 \text{ mm}^{-1}$ and $\mu_s = 6.86 \text{ mm}^{-1}$. The scattering media had a width of 12 mm and a depth of 6 mm. The center of the upper side was defined as the origin, and the horizontal right and vertical downward directions were set as the $X$ and $Y$ axes, respectively. The width of the scattering media is sufficiently large for optical scattering in the horizontal direction. Right-handed CPL ($S_3 = +1$) beams were irradiated with $(x, y) = (0, 0)$. The incident angle $\theta$ were fixed at $\theta = 1°$ and the detection angles $\varphi$ were defined as angles from the normal of the surface of the turbid media. In the reflection configuration, the detector faces a region of 2 mm width from the point at a distance of 1 mm from the origin, that is, from $(1, 0)$ to $(3, 0)$. In the transparent configurations, the

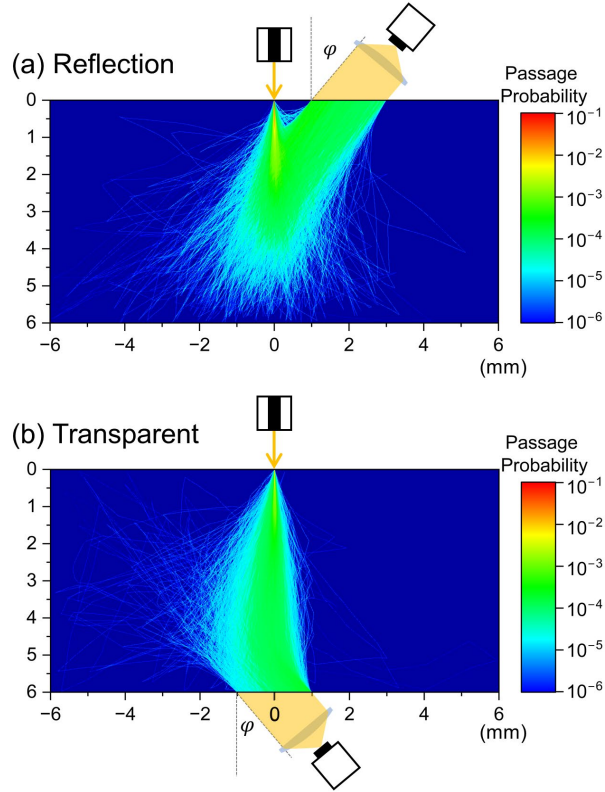

**Figure 1** (a) Reflection and (b) transparent configurations for multiple scattering events in pseudo-turbid medium, together with the distribution of the passage probability of light beam under the condition that the detection angle $\varphi = 40°$, $a = 10$ μm, and $\lambda = 1000$ nm in the medium with the lateral 12 mm ($x$) and the longitudinal 6 mm ($y$) in dimensions. CPL beams are irradiated into the medium at $(x, y) = (0, 0)$ point. The detector is located on (a) the same side and (b) the opposite side of the CPL source, which collect the scattered light beams emitted from the region (a) from $(1, 0)$ to $(3, 0)$ and (b) from $(-1, 6)$ to $(+1, 6)$, respectively.

detector is located at the opposite side of the scattering media, which can detect the light emitted from the region from $(-1, 6)$ to $(+1, 6)$. The distributions of the passage probability of a light beam are depicted in the scattering medium region in Figure 1, which were calculated under the condition that the detection angle $\varphi = 40°$, $a = 10$ μm, and $\lambda = 1000$ nm. The wavelength of incident CPL beams varied from 600 to 1500 nm, and the ratio of diameter and wavelength defined by $X \equiv a/\lambda \ (= x/\pi \cdot n_{medium})$, which is called "generalized size parameter", was used to compare their contributions to depolarization. Except for the calculations for refractive index dependence, the refractive indices of a particle and the matrix are fixed at 1.59 and 1.33, which are typical values for biological materials and water, respectively [12, 13]. In this case, the relative refractive index $m$ is $m \cong 1.195$.

## 4 Results

### 4.1 Single Scattering

Figure 2 shows the representative calculation results for a single scattering event in the case of $\lambda = 1000$ nm and $a = 9.0$ and $11.0$ μm corresponding to $X = 9.0$ and $11.0$, respectively. The incident light propagates from left ($\theta = 180°$) and collides with a particle at the origin. Figure 2 (a) shows the angular distribution of the relative intensity of the scattered light on polar coordinates with the radial axis of the logarithmic scale, which is equivalent to the scattering probability. The angular scattering distributions are dominated by forward scattering ($\theta = 0°$ direction) and include fine interference fringes in the other directions resulting from the reradiation of the excited dipoles. The differences in particle size cause some variations in occupancy of the straight-forward direction ($\theta = 0°$) for all other directions. The intensity distribution for $X = 9.0$ indicates larger back and side scattering instead of a slightly smaller straightforward scattering compared with that for $X = 11.0$. Similar behavior with violent vibrations was observed in the angular distribution of the resultant circular polarization (Figure 2(b)). The light scattered toward the straight-forward direction ($\theta = 0°$) maintains its circular polarization ($S_3 = +1$) and the helicity of the straight-back scattered light ($\theta = 180°$) is turned over ($S_3 = -1$). The $S_3$ distribution definitely passes through these two points, $(\theta, S_3) = (0°, +1)$ and $(180°, -1)$ irrespective of $a$ and $\lambda$. In the Rayleigh regime, the cosine curves connect these two points to the origin [12]. In contrast, in the Mie regime, the light scattered at intermediate angle between them undergoes depolarization and interference, resulting in the complicated and oscillatory changes.

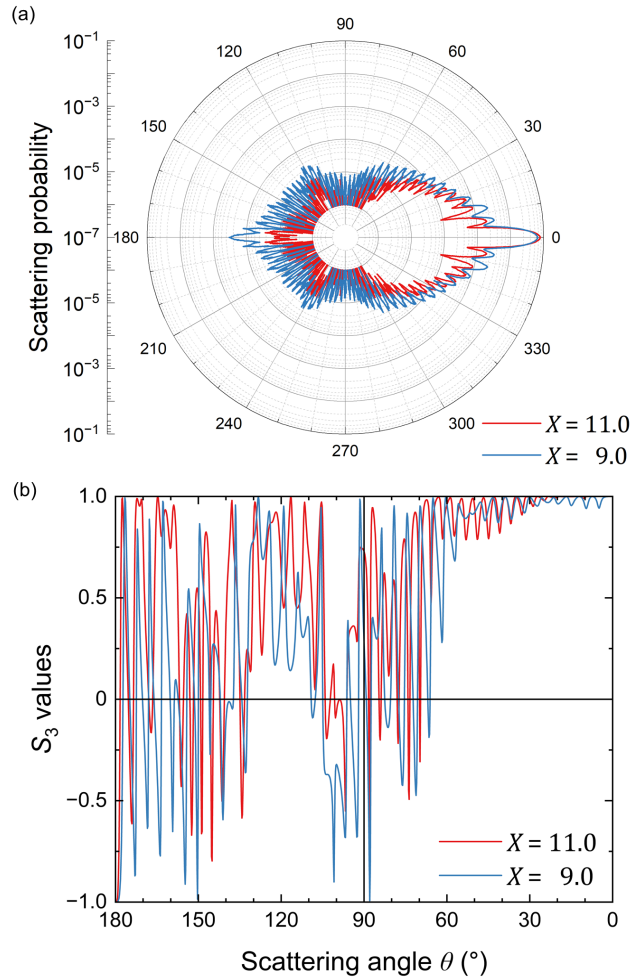

**Figure 2** Scattering angle $\theta$ dependence of (a) intensity and (b) $S_3$ values of scattered light for a single scattering event for (red) $X = 11.0$ and (blue) $X = 9.0$. A right-handed CPL ($S_3 = +1$) is incident from the direction $\theta = 180°$ toward a particle located at the origin.

*4.2 Size Parameter Dependence of Depolarization*

The expected value of the $S_3$ value depolarized by a single scattering event can be provided by the sum of the products of $S_3$ and the scattering probability at each angle. The black lines and plots in Figure 3 represent the expected values of the $S_3$ as a function of $X$, ranging from $X = 1$ to $30$ on the left axis. At specific $X$ values, such as $X \cong 5, 9, 12, 17, 19, 22,$ and $26$, the scattered light is subjected to comparatively large depolarization. By introducing the calculated angular distributions of $S_3$ and the scattering probability into the MC algorithm for multiple scattering, the expected value of the circular polarization of light scattered from turbid media can be obtained. The resultant polarization values of the scattered light in the reflection and transparent configurations are represented by the red and blue lines and plots in Figure 3, respectively. Because the depolarization effects are accumulated by hundreds of scattering events, the circular polarization of the scattered light emitted outward in both configurations inherits the $X$-dependent characteristics of single scattering. Particularly at the points $X \cong 9, 12,$ and $22$ among the specific $X$ values where large depolarization in single scattering is obtained, a drastic decrease in $S_3$ value emerges cumulatively. The transmitted scattered light exhibits weaker depolarization than that in the reflected geometry over almost the entire $X$ region. This is because most of the light detected on the opposite side of the light source experiences more straightforward scattering. The scattered light detected on the same side undergoes many

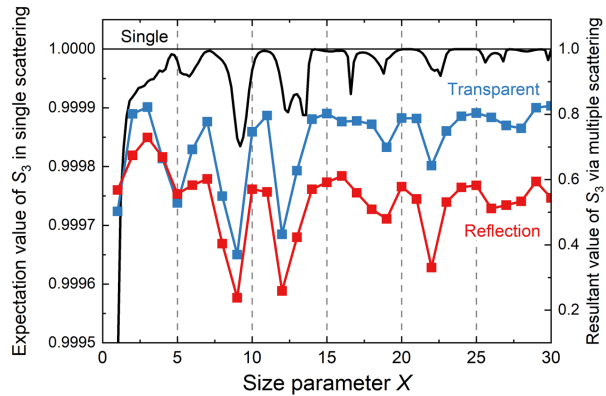

**Figure 3** Size parameter dependence of (black) the expected value of the $S_3$ value depolarized by a single scattering event on the left axis, and the resultant $S_3$ values of multiple scattering on the right axis in (red) reflection and (blue) transparent geometries. Resultant $S_3$ values are obtained as the averages over the scattered angle $\varphi$.

scattering events with bending in that direction, causing larger loss of polarization.

*4.3 Depolarization Diagrams*

The size parameter $X$ dependences of depolarization shown in the preceding paragraph are extended to wavelength $\lambda$ dependence, yielding the depolarization diagrams with particle size and wavelength as independent variables. Figure 4 shows the depolarization diagram of scattered light in (a) reflection and (b) transparent configurations, using the diameter of the particle $a$ (μm) as a horizontal axis and the wavelength $\lambda$ (nm) as a vertical axis. In these diagrams, larger reddish areas indicate weaker depolarization, and larger bluish areas display stronger depolarization. In this study, the blue linear region connecting $(a, \lambda) = (5.4, 600)$ with $(13.5, 1500)$ is called the $X = 9$ line. Similarly, the line-shaped regions connecting $(7.2, 600)$ with $(18.2, 1500)$ and $(13.2, 600)$ with $(33.0, 1500)$ are called the $X = 12$ and $X = 22$ lines, respectively. These line-like regions show intensive depolarization, which commonly appears in reflection and transparent geometries. In the reflection configuration, in the smaller $X$ regions on the left side of the $X = 9$ line, there was almost no change in the depolarization intensity, whereas there were slightly strong depolarization

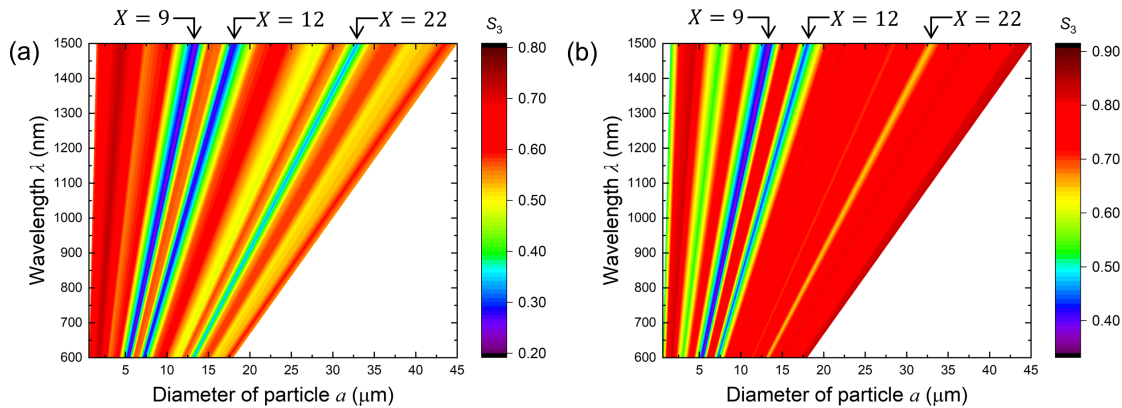

**Figure 4** Depolarization distribution diagram for (a) reflection and (b) transparent geometries. Three lines appearing in both diagrams indicate intensive depolarization, $X = 9, 12,$ and $22$ lines

(yellowish) areas corresponding to the $X = 19$ and $X = 26$ on both sides of the $X = 22$ line. On the contrary, in the transparent geometry, the $X = 22$ line is narrow and weak, and wide constant areas extend in its vicinities, whereas the wide and slightly stronger depolarization region ($X \cong 5$) is extended linearly in the smaller $X$ region. These characteristics were derived by integrating the difference in scattering angles over the scattering events.

*4.4 Refractive Indices Dependence*

In the preceding calculations, the refractive indices of a particle and the matrix are fixed at 1.59 and 1.33, yielding $m \cong 1.195$. Figure 5 shows the expected value of the $S_3$ for a single scattering as a function of $X$ on the horizontal axis and the relative refractive index $m$ on the vertical axis. In the region of $m \gtrsim 1.2$, the depolarization distribution shows the complicated but periodical moiré patterns, whereas the nearly constant depolarization is shown in $1 < m \leqq 1.2$. Assuming an organic spherical component in a biological liquid, the $m$ value is larger than 1 and close to 1. For example, the refractive index of cell nucleus and the environmental liquid (cytoplasm of the cell) is approximately 1.39 and 1.37, respectively [24, 25], $m = 1.015$. Therefore, the contribution of $m$ value to depolarization can be ignored in biological applications. However, high refractive index particles require examination of the $m$ contribution. The result shown in Figure 5 is compatible with the results calculated by Macdonald *et al* [9].

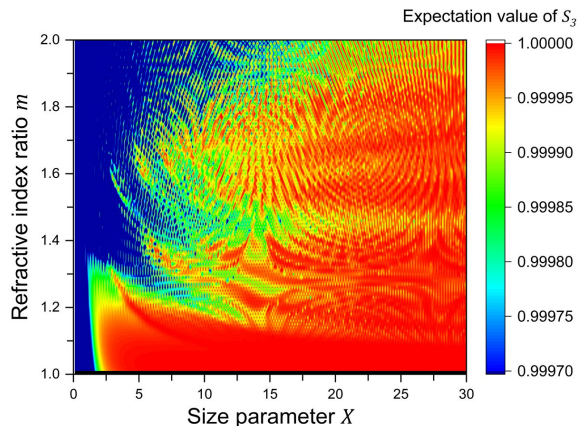

**Figure 5** The expected value of the $S_3$ for a single scattering as a function of $X$ on the horizontal axis and the relative refractive index $m$ on the vertical axis. Since the scattering event rarely occurs near $m = 1$, the data is omitted.

## 5 Discussion

The depolarization diagrams shown in Figure 4 assist us in examining the feasibility of the measurement technique for various particles in turbid media and determining the optimum wavelengths. Consider the detection of cancer distributions previously stated based on the diagram of the reflection configuration (Figure 6). The cell nuclei in normal tissues of the stomach wall are approximately 6 μm in diameter. Draw a vertical line at $a = 6$ μm on the diagram, then it crosses the $X = 9$ line at $\lambda \cong 670$ nm, which indicates that CPL of 670 nm in wavelengths is depolarized strongly in normal tissues. While the tissue becomes cancerous, the cell nuclei grow larger, approximately 11 μm on average. A vertical line at $a = 11$ μm intersects with the $X = 12$ and $X = 9$ line at $\lambda \cong 920$ and $\lambda \cong 1240$ nm, respectively. The diagram shows weak depolarization (reddish) near the intersection points of horizontal lines of these wavelengths and $a = 6$ μm. Therefore, the CPL of these two wavelengths suffers from drastic depolarization only for scattering with cancerous cell nuclei. In contrast, the horizontal line of $\lambda = 670$ nm passes the reddish area near $a = 11$ μm, implying that the cancerous tissues cause no remarkable depolarization to the light with $\lambda = 670$ nm in wavelength. Considered together, cancerous cells can be quantitatively detected in contrast to normal cells using light with a wavelength of 670 nm and one or both wavelengths of 920 and 1240 nm [18].

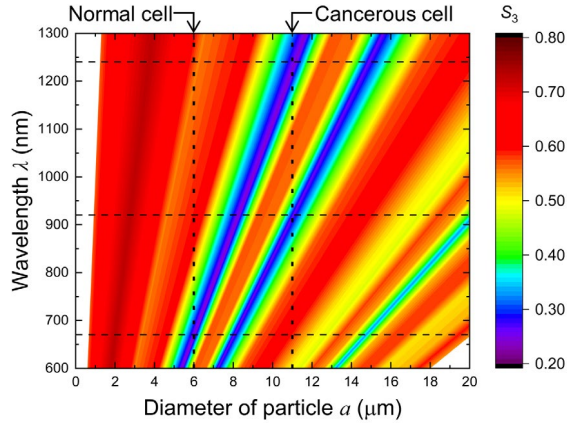

**Figure 6** Wavelength optimization for cancer detection using the depolarization diagram for the reflection geometry (Figure 4 (a)). The averaged diameter of the normal cell nucleus, $a = 6$ μm, and that of the cancerous cell nucleus, $a = 11$ μm, are denoted by the vertical dotted lines, which intersect $X = 9$ line and $X = 12$ and $9$ lines at $\lambda = 670$ nm, and $\lambda = 920$ and $1240$ nm, respectively.

The MC calculations in this study did not consider the variations of absorption and scattering coefficients. These coefficients, which give us the mean free path length and the optical attenuation during scattering events, contribute to the number of scattering events, but not affect the relationship between wavelength and the size of particle in a single scattering. In this paper, we focus the contributions of difference in particle size on the circular polarization of scattered light by fixing the number of scattering events. These optical coefficients vary owing to the tissue and the environments, accordingly, the number of scattering events changes. However, extensive feasibility studies of the respective applications, including the parameters corresponding to the application target, should be reviewed with appropriate coefficients for each application. For cancer detection, cancerous tissues have the comparatively larger optical coefficients than that of normal tissues. However, we confirmed that the changes resulting from the changes of particle size is significantly large for the changes from the number of scattering events [18]. Therefore, a light with a wavelength of 1240 nm is unsuitable because of its strong optical absorption by water. For biological applications with visible and infrared light, the spectral optical parameters for target tissues can be approximated as a weighted sum of the three constituent optical parameters for oxy-hemoglobin, deoxy-hemoglobin, and water [26–28]. In addition, the choice between reflection and transparent configurations should also be determined by the form and environment of the objects of interest, as well as whether mild depolarization is acceptable, such as the $X = 12$ and $X = 22$ in reflection and $X \cong 5$ in transparent geometry.

Optical scattering is stochastic phenomenon. However, the integration over a sufficient number of scattering events or taking for a sufficiently long time constant, provides stable values of circular polarization. In a thin tissue or a low-density turbid, the number of scattering events is not secured. And, the time constant to obtain a stable polarization values depends on size stability of scatterers, flow speed of scatterers and changing rate in surface roughness. For blood-cell monitoring, the time constant should be sufficiently longer than the pulse period. Therefore, the sampling rate is low. In case of monitoring of almost non-movable scatterers the shorter time constant is sufficient. Macdonald *et al* investigated the survivals of CPL depolarization in polydisperse scattering media [9]. They reported that the number of scattering events required for complete depolarization depends strongly on the size parameter and refractive index ratio. The time constant appropriate for respective applications can be found from their calculations. Temporal changes of surface profile have a great influence on polarization state of scattered light for any applications. In case of *in vivo* diagnosis of stomach cancer, the influences of gastric peristalsis and secretion of bodily fluid (gastric acid) should be suppressed. The other factors such as temperature and pH have no significant effect, unless these affect the properties of scatterers and surface of measurement points.

## 6   Proposals

When measuring particles with one diameter in turbid media, including particles of different sizes, two or more wavelengths in which intensive depolarization occurs only for the particles of interest and the other particles in the matrix should be selected based on the depolarization diagrams. Several applications based on this procedure are proposed in this section. Note again that the feasibility of the respective applications needs to be examined with the optical parameters, the diameter dispersion of particles, and configurations appropriate for the measurement target, condition and environment. The narrow dispersion of particle sizes about one dominant size has the effect of smoothing out the short scale fluctuations shown in figure 3 [9]. Biological tissues include various particles and structures having sizes largely deviated from dominant ones. And, as posed by Borovkkova *et al.*[29], the contributions on depolarization of birefringence due to the connective interface in biological tissues should be taken into consideration. Such complexity of biological tissues would cause a uniform depolarization; however, these contributions should be reviewed by further calculations and experiments.

*6.1 Blood Cell Monitoring*

White blood cells (WBCs) are corpuscles that protect the body from pathogens such as bacteria and viruses, and their status is used as an indicator of immunological status in the diagnosis and treatment of various medical conditions, such as infectious diseases [30, 31], sepsis [32], cancers [33], and autoimmune disorders [34]. The normal WBC count is typically between 4,000 and 11,000 per 1 microliter of blood, which is 1/1000 of the red blood cell (RBC) count. The diameter of WBC is $12-15$ μm, whereas that of RBC is $6-8$ μm. We considered the feasibility of noninvasive WBC counting using CPL scattering based on the depolarization diagram for the transparent configuration, assuming the measurement at a tip of a finger or an ear lobe. In this study, we did not consider the internal structure and cell lineage [35], but the size of the blood corpuscles. Figure 7 shows the extract from Figure 4 (b). The areas interposed between the vertical dotted lines represent the regions of the WBC and RBC sizes, respectively, and the dotted horizontal lines are drawn on the intersections of these vertical lines and the intensive depolarization area ($X = 9$ and $11$ lines) for eye guides. The thick black frames indicate areas where large depolarization lines were included within the width of the corpuscle. Light with a wavelength ranging from 1000 to 1250 nm causes depolarization of WBCs rather than RBCs. In contrast, the depolarization of CPL at wavelengths ranging from 670 to 900 nm is derived only from RBC. These wavelengths are provided as the upper and lower sides of the frames. A comparison of the depolarizations between these wavelength regions provides a relative change in the WBC count. Such a method will result in noninvasive and simple blood cell monitoring in real time [36, 37], which will reduce the mental and physical burdens of drawing blood from patients, particularly, immature infants, leukemic children, or massive bleeding patients. Moreover, this technique will contribute to the investigation of the immune system of various living organisms, that is, morphological studies on fish diseases in aquaculture [38] and health monitoring of domestic animals.

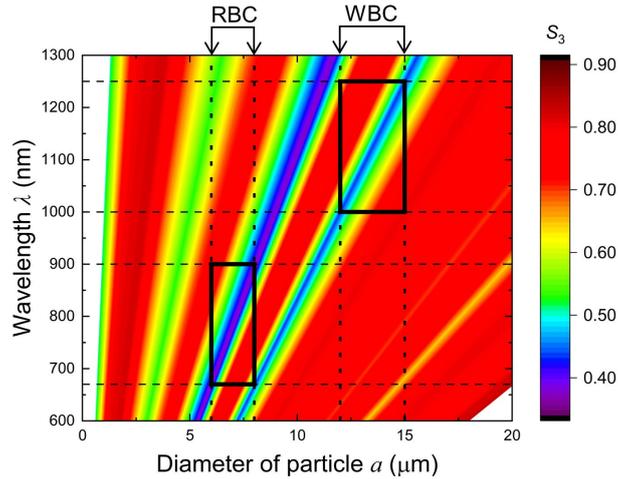

**Figure 7** Wavelength optimization for blood cell monitoring using the depolarization diagram for the transparent geometry (Figure 4 (b)). The dispersion width of the diameter for RBC, $a = 6 - 8$ μm and WBC, $12 - 15$ μm, are denoted by the vertical dotted lines. These lines make rectangles delineated by thick black lines whose diagonal lines are the lines between the intersections with $X = 9$ line and $X = 12$ lines, respectively. The lower and upper sides of the rectangles are on the horizontal lines, $\lambda = 670$ and $900$ nm, and $\lambda = 1000$ and $1250$ nm, respectively.

*6.2 Early Embryogenesis Process*

Early embryonic cells undergo frequent division. Rapid cell division increases the number of cells without intervening in cell growth over approximately four days, progressing first to the morula stage and then to the blastocyst stage. Over the course of pre-implantation development, the size of the cell nucleus decreases to one-tenth of its original size [39]. Assuming that the nucleus is spherical, the diameter of nuclei progressively scales down from approximately $25\ \mu m$ for the male pronucleus in 1-cell stage embryos to approximately $11\ \mu m$ for the nuclei in blastocysts. The depolarization diagrams shown in Figure 4 suggest that the stepwise decrease in the nucleus diameter can be followed through the spectrum of the scattered CPL over a wide wavelength range of $600 - 1150\ nm$. The $S_3$ spectrum of the scattered light showed sharp depolarization dips at wavelengths corresponding to the nuclear diameter. The nucleus size can be estimated based on the $X = 22$ depolarization line from approximately $25\ \mu m$ (1-cell) to approximately $15\ \mu m$ (8-cell). For the further divisions, depolarization can be observed in the $S_3$ spectrum in the long- and short-wavelength regions of $900 - 1500\ nm$ and $600 - 700\ nm$, respectively. Observation and control of the cleavage processes of early embryos will become increasingly important as practical research on stem cells advances, such as embryonic stem (ES) cells [40] and induced pluripotent stem (iPS) cells [41]. This optical and noninvasive observation (counting) technique using CPL scattering will be useful not only for embryology and developmental biology but also the stem cell research.

*6.3 Antigen-antibody interaction*

Antibodies are proteins that play a leading role in humoral immunity induced in the living body because of the immunoreaction caused by antigen stimulation. The antigen-antibody interaction, in which an antigen and the corresponding antibody are conjugated, has extremely high specificity and a high molecular recognition function. The amount of objective protein can be measured by chemically linking antibodies or antigens with a detectable label, which either changes the color of the solution, fluoresces under ultraviolet light, or emits light [42]. We now consider polymer particles that are surface-modified with an antibody. The connection to a target protein or antigen increases the apparent particle diameter. Depending on the particle size before and after bonding, CPL scattering is expected to be a label-free immunoassay, which can be applied to clinical examinations with low invasion.

# 7  Conclusions

Circularly polarization of light survives more scattering events than linear polarization in the Mie scattering regime, where the scatterer size is comparable to the wavelength of the light. Moreover, the depolarization of circularly polarized light owing to particle scattering is strongly dependent on the size parameters $X$, which are the ratio between the wavelength and particle size. The depolarization intensity provides structural information about the particles, such as their size, density, and distribution. The correlation between depolarization intensity, wavelength, and particle size was examined using MC simulation method. In single-scattering, intensive depolarization occurs for specific size parameters. Calculations for multiple scattering in reflection and transparent geometries were performed by introducing single-scattering behavior into the MC algorithm. Consequently, the characteristic features of depolarization for a single scattering were emphasized for some $X$ values. By independently changing wavelength and particle diameter, depolarization distribution diagrams using these two parameters as orthogonal axes were obtained for the reflection and transparent geometries. When investigating the feasibility of various particle monitoring methods, these diagrams are a particularly useful guide. In addition to cancer-detection techniques using CPL scattering, several applications based on these diagrams have been proposed. Large differences in size between WBC and RBC have been used in blood cell monitoring. In the depolarization diagrams, the wavelengths at which intensive depolarization occurred for each corpuscle size were selected. Early embryos develop rapidly by dividing cells, and the size of the cell nucleus decreases stepwise over the first few days. Stepwise changes in the nucleus size can be observed using broadband CPL, which can cover the wavelength regions corresponding to the initial to the final target size of the nucleus. Moreover, the surface-modified polymer particles apparently increase in size owing to adsorption of proteins or viruses via antigen-antibody interactions, which can be detected using CPL at the appropriate wavelength. In addition, various applications of CPL scattering can be proposed, such as observations of the incorporation process of graft tissue and monitoring the state of roe in fish without laparotomy. Note that the feasibility of these proposed applications needs to be reviewed with the optical parameters, contribution of birefringence and configurations appropriate for the particle condition and measurement environment. The further study would expand the applicable range of this CPL scattering technique to a broader range of biological conditions and tissues, which would develop the versatility of the methodology.

Optical observation methods for biological materials are characterized by noninvasive and rapid responses, which are required in various fields. Several techniques based on various principles have been developed and used to provide useful information from various perspectives. If research into the CPL scattering method progresses, it would add another aspect to their understanding.

**Ethics approval**

Ethics approval is not required to carry out this study.

**Disclosures**

The authors declare that there are no conflict of interest related to this article.

**Acknowledgments**

This study was partially supported by KAKENHI (Nos. 19H04441 and 22H03921) of the Japan Society for the Promotion of Science (JSPS), the Cooperative Research Project of the Research Center for Biomedical Engineering, the Futaba Foundation, the Uehara Memorial Foundation, and grants from the Kitasato university (No. 2023-1768). The authors acknowledge Dr. Masamichi Oh-Ishi, Mr. Mike Raj Maskey and Mr. Yutaro Utsumi for fruitful discussion and technical support at Kitasato University.

**Code and Data Availability**

Data and Code supporting the findings of the article are publicly available, and can be obtained from the authors upon reasonable request at nishizawa.nozomi@kitasato-u.ac.jp.